\documentclass[twocolumn]{aastex631}
\usepackage{amsmath}
\usepackage{graphicx}
\usepackage{rotating}
\usepackage{CJK}

\begin{document}
\begin{CJK*}{UTF8}{gbsn}
\title{On the occurrence of stellar fission in binary-driven hypernovae}


\author[0000-0003-0368-384X]{S. R. Zhang (张书瑞)}

\affiliation{ICRANet, Piazza della Repubblica 10, I-65122 Pescara, Italy}
\affiliation{School of Astronomy and Space Science, University of Science and Technology of China, Hefei 230026, China}
\affiliation{CAS Key Laboratory for Research in Galaxies and Cosmology, Department of Astronomy, University of Science and Technology of China, Hefei 230026, China}

\correspondingauthor{S. R. Zhang}
\email{zhangsr@mail.ustc.edu.cn}

\author[0000-0003-0829-8318]{R. Ruffini}
\affiliation{ICRANet, Piazza della Repubblica 10, I-65122 Pescara, Italy}
\affiliation{ICRA, Dipartimento di Fisica, Sapienza Universit\`a di Roma, P.le Aldo Moro 5, I-00185 Rome, Italy}
\affiliation{INAF, Viale del Parco Mellini 84, I-00136 Rome, Italy}
\email{ruffini@icra.it}






\begin{abstract}
The binary-driven hypernova (BdHN) model address long gamma-ray bursts (GRBs) associated with type Ic supernovae (SNe) through a series of physical episodes that occur in a binary composed of a carbon-oxygen (CO) star (of mass $\sim 10 M_\odot$) and a neutron star (NS) companion (of mass $\sim 2 M_\odot$) in a compact orbit. The SN explosion of the CO star triggers sequence of seven events. The BdHN model has followed the traditional picture of the SN from the CO iron's core collapse. However, the lack of a solution to the problem of producing successful SNe leaves room for alternative scenarios. We here show that tidal synchronization of the CO-NS binary can lead the CO star to critical conditions for fission, hence splitting into two stellar remnants, e.g., $\sim 8.5 M_\odot +1.5 M_\odot$. We give specific examples of the properties of the products for various orbital periods relevant to BdHNe. The astrophysical consequences of this scenario are outlined. 

\end{abstract}




\section{Introduction} \label{sec:1}

The origin of long gamma-ray bursts (GRBs) is thought to be related to the death of massive stars, being their association with type Ic supernovae (SNe) \cite{1998Natur.395..670G, 2006ARA&A..44..507W, 2011IJMPD..20.1745D, 2012grb..book..169H} one of the most compiling observational evidence. Since most massive stars belong to binaries \citep[see, e.g.,][]{2007ApJ...670..747K, 2012Sci...337..444S}, the direct collapse of a massive star to a BH should not produce an SN, observed pre-SN progenitors have masses $\lesssim 18~M_\odot$ \citep{2009ARA&A..47...63S, 2015PASA...32...16S}, and stellar evolution models predict the direct formation of a BH only in progenitor stars $\gtrsim 25 M_\odot$ \citep[see, e.g.,][]{2003ApJ...591..288H}, one can conclude that the GRB and the SN should not originate from a single star. Based on the above theoretical and observational clues, the binary-driven hypernova (BdHN) model proposes a binary system composed of a carbon-oxygen star (CO) and a neutron star (NS) companion as the progenitor of GRB-SNe. We refer the reader to \citet{2012ApJ...758L...7R,2014ApJ...793L..36F,2015PhRvL.115w1102F,2016ApJ...833..107B,2018ApJ...852...53R,2018ApJ...869..101R,2019ApJ...871...14B,2019ApJ...886...82R,2020EPJC...80..300R,2021A&A...649A..75M,2021MNRAS.504.5301R,2022ApJ...929...56R,2022ApJ...939...62R,2022ApJ...936..190W,2022PhRvD.106h3004R,2022PhRvD.106h3002B}, for theoretical details and applications of the BdHN model to specific sources, and to \citet{2023arXiv230316902A} for the latest developments. 

The BdHN model assumes the occurrence of the SN onsets the entire cataclysmic event: the core collapse of the CO forms a newborn NS (hereafter $\nu$NS) and ejects material that accretes onto the NS companion and the $\nu$NS owing to matter fallback. In this picture, the binary's orbital period is a critical parameter determining the system's fate and energetics, which leads to the classification of BdHNe into type I ($\gtrsim 10^{52}$ erg), II ($\sim 10^{50}$--$10^{52}$ erg), and III ($\lesssim 10^{50}$ erg; see \citealp{2023arXiv230316902A}, for details).

\begin{figure}
\centering\includegraphics[width=\hsize,clip]{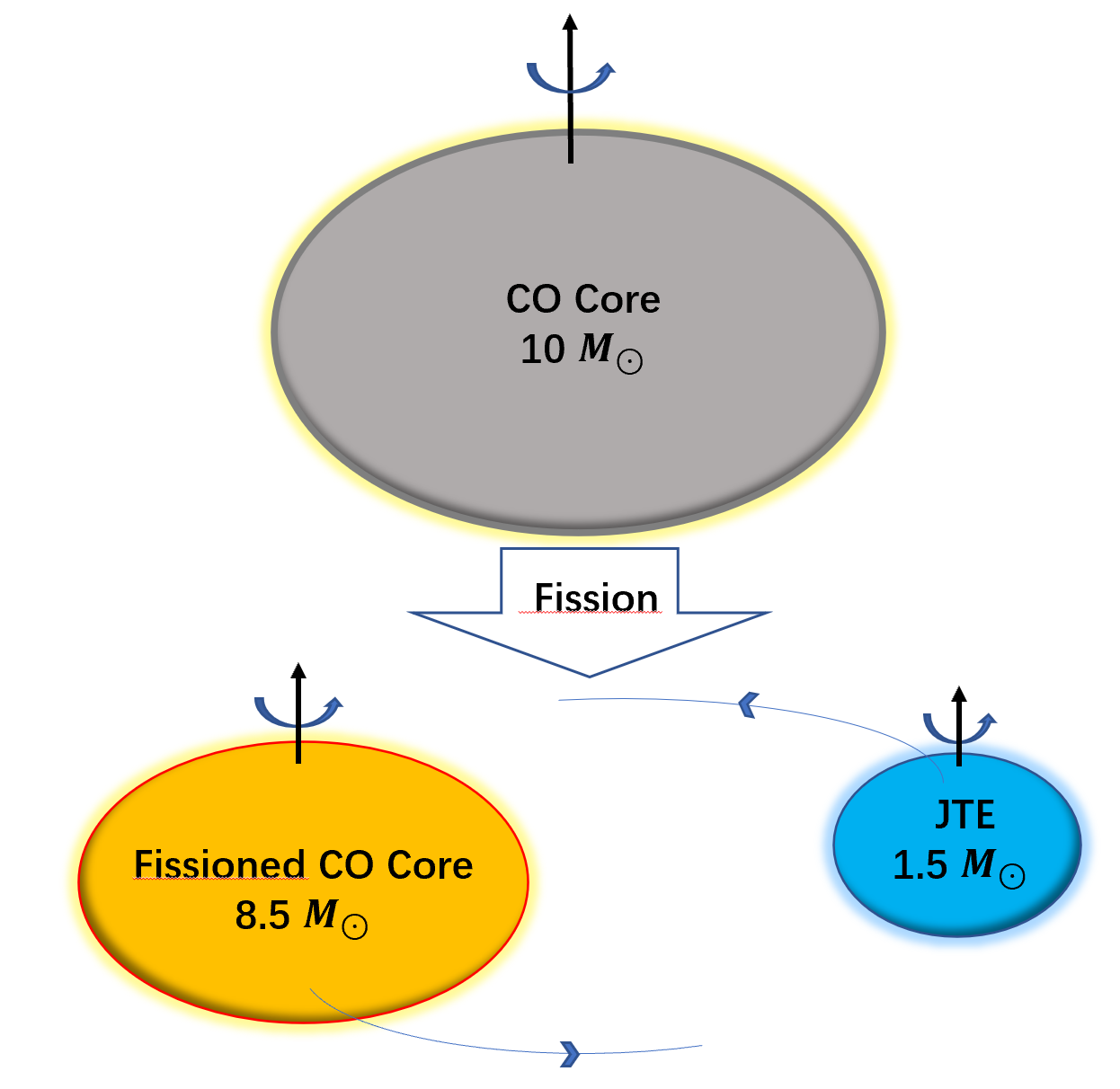}
\caption{Schematic diagram illustrating the fission of the CO star into the fission CO core and JTE.}
\label{fig:betaless30}
\end{figure}

Since the SN trigger and the $\nu$NS formation are crucial in explaining the event, a vast new topic has emerged from studying alternatives to assuming a single non-rotating star core-collapse SN event in the CO core. The alternative scenario has been recently advanced based on the fission process of the CO core due to its high rotation rate gained by corotation with its NS companion. It has been indicated in \citet{2023arXiv230316902A} that the GRB-SN might be triggered, e.g., by a fast rotating $10 M_{\odot}$ CO star set in corotation with a companion NS in an orbital period of a few minutes. The CO fission creates an $8.5 M_{\odot}$ Maclaurin ellipsoid core and a $1.5 M_{\odot}$ companion triaxial Jacobi ellipsoid (hereafter, JTE) in a Roche-lobe configuration (see Fig. \ref{fig:betaless30}). 

The aim of this article is to evaluate specific examples of the above fission process. For this task, we have generalized (see appendix \ref{sec:app}) the classical tables of the Maclaurin and Jacobi sequences in \citet{1969efe..book.....C, jeans1929astronomy} treatises necessary to describe the CO core fission in the present astrophysical scenario. 

In Section \ref{sec:3}, we illustrate a few examples of the binary progenitor before fission, i.e., an initial $10 M_{\odot}$ CO core corotating with an NS companion for selected values of the orbital period. Section \ref{sec:4} shows examples of the system after the fission of the initial $10 M_{\odot}$ CO core into a Maclaurin new CO core of $8.5 M_{\odot}$ and a JTE companion of $1.5 M_{\odot}$ (see Fig. \ref{fig:betaless30}). Mass and angular momentum conservation are necessary conditions in the fission process.

To exemplify, we assume the CO core after fission lies on a specific location of the Maclaurin sequence and the JTE, by definition, on the Jacobi sequence. One could select alternative locations for the fission products, and the outcomes would not vary significantly, as the physical quantities are similar. The five examples of fission we are considering are displayed in Tables \ref{tab:table3} and \ref{tab:table4}. Table \ref{tab:table3} lists the physical properties of the CO before fission, while Table \ref{tab:table4} presents the physical properties of the fission products. The initial CO core in all examples has a mass of $10 M_{\odot}$, while the rotation period varies.

\section{Before fission} \label{sec:3}

\begin{table*}[htbp!]
\centering
\caption{\label{tab:table3}%
The physical quantities before fission. The mass of the CO core is fixed at $10 M_{\odot}$. Unless otherwise specified, we use cgs units. From left to right, the table's columns are the spin period, density, average radius, angular momentum, rotational energy, and moment of inertia of the CO core.}
\begin{tabular}{cccccc}
\hline
$P$ & $\rho$ & $\bar{a} $ & $J_{\rm CO}$ & $E_k$ & $I$\\
(min) & (g cm$^{-3}$) & (cm) & (g cm$^2$ s$^{-1}$) & (erg) & (g cm$^2$) \\
\hline
110 & 11.5573 & $7.43424\times 10^{10}$ & $6.00045\times 10^{52}$ & $2.85621\times 10^{49}$ & $1.09934\times 10^{56}$ \\
\hline
90 & 17.2647 & $6.50334\times 10^{10}$ & $5.61221\times 10^{52}$ & $3.26505\times 10^{49}$ & $8.4126\times 10^{55}$ \\
\hline
40 & 87.4023 & $3.78746\times 10^{10}$ & $4.28292\times 10^{52}$ & $5.60633\times 10^{49}$ & $2.85334\times 10^{55}$ \\
\hline
20 & 349.609 & $2.38595\times 10^{10}$ & $3.39935\times 10^{52}$ & $8.89949\times 10^{49}$ & $1.13235\times 10^{55}$ \\
\hline
5 & 5593.75 & $9.46866\times 10^9$ & $2.14146\times 10^{52}$ & $2.24253\times 10^{50}$ & $1.78334\times 10^{54}$ \\
\hline
\end{tabular}
\end{table*}

Let us assume that the binary consists of a CO core of mass $M_{\rm CO}$ and an NS of mass $M_{\rm NS}$. The corotation implies the CO's core spin period is the same as the orbit period, $P$. We denote the rotational angular velocity of the CO core as $\Omega$. The initial CO core is assumed to be at the Maclaurin-Jacobi bifurcation point, at which the angular velocity and density fulfill \citep[see, e.g.,][]{1969efe..book.....C}
\begin{equation}\label{eq:bifurcation}
    \frac{\Omega^2}{\pi G \rho_{\rm CO}} \approx 0.37423.
\end{equation}
The rotation angular momentum of the CO core is 
\begin{equation}\label{eq:J}
    J_{\rm CO} = I_{\rm CO}\,\Omega,
\end{equation}
where the moment of inertia is given by
\begin{equation}\label{eq:I}
    I_{\rm CO} = \frac{2}{5} M_{\rm CO}\,a_{\rm CO}^2,\quad a_{\rm CO} = \bar{a}_{\rm CO} (1-e^2)^{-1/6},
\end{equation}
with the reference radius $\bar{a}$ (of the spherical configuration with the same volume) set by 
\begin{equation}\label{eq:a0}
    \bar{a}_{\rm CO} = \left(\frac{3 M_{\rm CO}}{4 \pi \rho_{\rm CO}}\right)^{1/3} \approx \left( \frac{3}{4} \frac{0.37423\,G  M_{\rm CO}}{\Omega^2} \right)^{1/3},
\end{equation}
where in the last expression, we used Eq. (\ref{eq:bifurcation}). Having defined all the above, the angular momentum (\ref{eq:J}) can be written as
\begin{equation}
J_{\rm CO} \approx 0.30375 \sqrt{G\,M_{\rm CO}^3\, \bar{a}_{\rm CO}},
\label{eq:LCO}
\end{equation}
where we used that the eccentricity is $e \approx 0.812670$ at the bifurcation point. The constants $0.37423$ in Eq. (\ref{eq:bifurcation}) and $0.30375$ in Eq. (\ref{eq:LCO}) are, respectively, the values of the dimensionless square of angular velocity and angular momentum at the Maclaurin-Jacobi bifurcation point, which correspond to the first entry in Tables \ref{tab:table1} and \ref{tab:table2}.

\section{After fission} \label{sec:4}

We turn now to the two systems after the fission process of the original CO core. The products of the process are a new CO core with a mass of $M_{\rm COf}$ and a JTE with a mass of $M_{\rm JTE}$. This process is shown in Fig. \ref{fig:betaless30}. In the fission process, the following conditions are assumed:

(i) The fission products corotate with the same period (Roche condition);

(ii) The new CO core is located at a given point of the Maclaurin sequence, and the second product is a JTE, so it lies on the Jacobi sequence (see examples provided below);

(iii) The mass is conserved, i.e., 
\begin{equation}\label{eq:Mconser}
    M_{\rm CO}=M_{\rm COf}+M_{\rm JTE}
\end{equation}

(iv) The angular momentum of the CO core is conserved, i.e.,
\begin{equation}\label{eq:Lconser}
J_{\rm CO}=J_{\rm CO f} + J_{\rm JTE} + J_{\rm orbit}.
\end{equation}
In analogy with Eq. (\ref{eq:LCO}), the angular momentum of the new CO core can be written as
\begin{equation}
J_{\rm COf} = I_{\rm COf}\, \Omega_{\rm f} = c_1\,\sqrt{G  \,M_{\rm COf}^3\, \bar{a}_{\rm COf}}, 
\end{equation}
where the reference spherical radius is
\begin{equation}\label{eq:a0f}
    \bar{a}_{\rm COf} = \left(\frac{3 M_{\rm COf}}{4 \pi \rho_{\rm COf}}\right)^{1/3} = \left( \frac{3}{4} \frac{c_2\,G  M_{\rm COf}}{\Omega_f^2} \right)^{1/3},
\end{equation}
and the constant $c_2$ is defined by
\begin{equation}\label{eq:Omegaf}
    \frac{\Omega_f^2}{\pi G \rho_{\rm COf}} = c_2.
\end{equation}

Likewise, we define
\begin{equation}\label{eq:Jjte}
J_{\rm JTE} = I_{\rm JTE}\, \Omega_{\rm f} = c_3\,\sqrt{G\, M_{\rm JTE}^3 \, \bar{a}_{\rm JTE}}, 
\end{equation}
where
\begin{equation}\label{eq:a0fjte}
    \bar{a}_{\rm JTE} = \left(\frac{3 M_{\rm CJTE}}{4 \pi \rho_{\rm JTE}}\right)^{1/3} = \left( \frac{3}{4} \frac{c_4\,G  M_{\rm JTE}}{\Omega_f^2} \right)^{1/3},
\end{equation}
and
\begin{equation}\label{eq:Omegafjte}
    \frac{\Omega_f^2}{\pi G \rho_{\rm JTE}} = c_4.
\end{equation}

The orbital angular momentum is given by
\begin{equation}
J_{\rm orbit}=  \Omega_f \left( M_{\rm COf}\, r_{\rm COf}^2 + M_{\rm JTE} \,r_{\rm JTE}^2 \right),
\label{eq:Lorbit}
\end{equation}
where 
\begin{equation}\label{eq:rs}
    r_{\rm COf}=\frac{M_{\rm JTE}}{M_{\rm CO}} r_{\rm inner}, \quad r_{\rm JTE}=\frac{M_{\rm COf}}{M_{\rm CO}} r_{\rm inner},
\end{equation}
are the components' distance to the center of mass, and
from Kepler's third law, we obtain their separation
\begin{equation}\label{eq:rinner}
    r_{\rm inner} = r_{\rm COf} + r_{\rm JTE} = \left(\frac{G M_{\rm CO}}{\Omega_f^2}\right)^{1/3}.
\end{equation}
In Eqs. (\ref{eq:rs}) and (\ref{eq:rinner}), we used the mass conservation equation (\ref{eq:Mconser}).

\begin{table*}[htbp!]
\centering
\caption{\label{tab:table4}%
Fission examples for products that are not at the bifurcation point. From left to right, the columns of the table are the spin period and mass of the initial CO core, the mass of the new CO core, the mass of JTE, the spin period of the fission products, the density of the new CO core and the JTE, the distance between two fission products, the average radius of the new CO core and the JTE.}
\begin{tabular}{cccccccccc}
\hline
$P$ & $M_{\rm CO}$ & $M_{\rm COf}$ & $M_{\rm JTE}$ & $P_f$ & $\rho_{\rm COf}$ & $\rho_{\rm JTE}$ & $r_{\rm inner}$ & $\bar{a}_{\rm COf}$ & $\bar{a}_{\rm JTE}$ \\
(min) & $(M_{\odot})$ & $(M_{\odot})$ & $(M_{\odot})$ & (min) & (g cm$^{-3}$) & (g cm$^{-3}$) & (cm) & (cm) & (cm)
\\
\hline
110 & 10 & $8.5$ & $1.5$ & 50.7241 & 56.0085 & 54.5032 & $6.77722\times 10^{10}$ & $4.16145\times 10^{10}$ & $2.35548\times 10^{10}$ \\
\hline
90 & 10 & $8.5$ & $1.5$ & 41.5016 & 83.6669 & 81.4182 & $5.92859\times 10^{10}$ & $3.64036\times 10^{10}$ & $2.06053\times 10^{10}$ \\
\hline
40 & 10 & $8.5$ & $1.5$ & 18.4452 & 423.562 & 412.178 & $3.45274\times 10^{10}$ & $2.1201\times 10^{10}$ & $1.20003\times 10^{10}$ \\
\hline
20 & 10 & $8.5$ & $1.5$ & 9.22255 & 1694.27 & 1648.73 & $2.17508\times 10^{10}$ & $1.33558\times 10^{10}$ & $7.55968\times 10^9$ \\
\hline
5 & 10 & $8.5$ & $1.5$ & 2.30565 & 27108. & 26379.4 & $8.63185\times 10^{9}$ & $5.30025\times 10^{9}$ & $3.00007\times 10^9$ \\
\hline
\end{tabular}
\end{table*}

Using Eq. (\ref{eq:Lconser}), given the CO core rotation period, its mass, the mass of the fission products, and their location on the equilibrium sequences, we can determine the angular velocity of the new binary system, $\Omega_{f}$, so the corresponding period, $P_{f}$. Subsequently, using Eq. (\ref{eq:rinner}), we obtain the separation of the fission products, $r_{\rm inner}$.  In the examples shown in Table \ref{tab:table4}, the new CO core is a Maclaurin spheroid of eccentricity $e=0.8$, so $c_1=0.29345$ and $c_2=0.36316$ (see Table \ref{tab:table4}), and the JTE has a semi-major axis ratio $a_2/a_1=0.92$, so $c_3=0.304602$ and $c_4=0.37319$ (see Table \ref{tab:table3}).

It is worth noting that $\Omega_{f}$ is greater than $\Omega$, so the angular velocity increases after the fission. In all selected examples, $P_{ f} \lesssim P/2 $. Additionally, all fission products exhibit higher density compared to the CO core before fission, i.e., $\rho < \rho_{\rm cof}$ and $\rho < \rho_{\rm JTE}$. Another important feature is that the inner binary separation is approximately equal to the sum of the average radii of the fissioned CO core and the JTE, denoted as $\bar{a}_{\rm COf}$ and $\bar{a}_{\rm JTE}$, respectively, i.e., $r_{\rm inner} \approx \bar{a}_{\rm COf} + \bar{a}_{\rm JTE}$.

\section{Discussion and conclusions}\label{sec:5}

In the context of the BdHN model of GRB-SNe, we have given examples of an alternative scenario to the core collapse of the CO star. The corotation with the NS companion leads the CO star to fission into a new CO core in the Maclaurin sequence and a JTE of smaller mass. We have given examples where an initial CO core of $10 M_\odot$ splits into a new Maclaurin spheroid CO core of $8.5 M_\odot$ and a JTE of $1.5 M_\odot$ (see Tables \ref{tab:table3} and \ref{tab:table4}).

The subsequent evolution of the new CO core in this scenario could potentially provide an alternative SN mechanism: during its collapse, the rotational and magnetic energies of the CO core increase, becoming sufficiently large to unbind the outermost stellar layers or even to disrupt the CO core (Rueda, Ruffini, and Zhang, in preparation).

The evolution of the JTE possibly into a $\nu$NS remains to be studied. If angular momentum is conserved in that process, the parameters of the systems presented here would lead to a $\nu$NS that explains the observations of the X-ray luminosity observed in the very early phases of three BdHNe: GRB 220101A at cosmological redshift $z=4.2$, GRB 090423 at $z=8.2$ and GRB 090420B at $z= 9.4$ (Rueda, Ruffini, and Zhang, in preparation; Carlo Bianco et al. (2023)). The transition of the $\nu$NS from the Jacobi to the Maclaurin sequence releases a large amount of energy in kHz gravitational waves that could be detectable by the Advanced LIGO and Virgo interferometers for sources located up to $\sim 100$ Mpc \citep[see][for details]{2022PhRvD.106h3004R}.

\begin{acknowledgments}
We are grateful for the fruitful discussions with Jorge Rueda, Bianco Bianco, Yu Wang, M. Della Valle, and Rahim Moradi in preparing the manuscript.
\end{acknowledgments}

\appendix

\section{Equilibrium properties along the Maclaurin and Jacobi sequences} \label{sec:app}

We start by generalizing the classical tables of the Maclarin and Jacobi sequences in \citet{1969efe..book.....C} (Table I and  Table IV) or \citet{jeans1929astronomy} (Table XVI and Table XVII) treatises by extending to physical quantities necessary to describe the process involving fission and evolution of the Maclaurin and Jacobi sequence that can explain the observations of the X-ray luminosity of BdHNe (Rueda, Ruffini, and Zhang, in preparation). The main differences lie in the physical quantities required to describe the dynamics of these systems. This includes the dimensionless expressions of the moment of inertia ($I$) and the rotational kinetic energy ($E_{k}$), both of which are functions of the rotational period ($P$). These quantities are crucial in formulating the overall energetics of the systems. For more detailed information, see Tables \ref{tab:table1} and \ref{tab:table2}.

\begin{table*}[b]
\centering
\caption{\label{tab:table1}%
Properties of the Jacobi ellipsoids. The columns from left to right include the following ratios and dimensionless quantities: the ratio of the semi-axes $a_2$ to $a_1$ (with rotation along the $a_3$ axis), the ratio of the semi-axes $a_3$ to $a_1$, the ratio of the average radius to the semi-axis $a_1$, the dimensionless square of the angular velocity, the dimensionless rotation period, the dimensionless angular momentum, the dimensionless moment of inertia, and the dimensionless rotational kinetic energy.}
\begin{tabular}{cccccccc}
\hline
$a_2/a_1$ & $a_3/a_1$ & $\bar{a}/a_1$ & $\Omega^2/(\pi G \rho)$ & $P/\left(\pi G^{-1} \rho^{-1}\right)^{\frac{1}{2}}$ & $L/\left(G M^3 \bar{a}\right)^{\frac{1}{2}}$ & $I/ \left( M \bar{a}^2 \right)$ & $E_{\rm k}/\left(G M^2 \bar{a}^{-1}\right)$ \\
\hline
1.00 & 0.582724 & 0.835259 & 0.374230 & 3.26934 & 0.303751 & 0.573347 & 0.080461 \\
\hline
0.96 & 0.570801 & 0.818311 & 0.373980 & 3.27044 & 0.303959 & 0.573932 & 0.080490 \\
\hline
0.92 & 0.558330 & 0.800866 & 0.373190 & 3.2739 & 0.304602 & 0.575754 & 0.080575 \\
\hline
0.88 & 0.545263 & 0.782882 & 0.371785 & 3.28008 & 0.305749 & 0.579013 & 0.080726 \\
\hline
0.84 & 0.531574 & 0.764330 & 0.369697 & 3.28933 & 0.307467 & 0.583909 & 0.080951 \\
\hline
0.80 & 0.517216 & 0.745168 & 0.366837 & 3.30212 & 0.309837 & 0.590699 & 0.081259 \\
\hline
0.76 & 0.502147 & 0.725351 & 0.363114 & 3.31901 & 0.312956 & 0.599696 & 0.081659 \\
\hline
0.72 & 0.486322 & 0.704832 & 0.358424 & 3.34065 & 0.316938 & 0.611287 & 0.082162 \\
\hline
0.68 & 0.469689 & 0.683554 & 0.352649 & 3.3679 & 0.321923 & 0.625965 & 0.082780 \\
\hline
0.64 & 0.452194 & 0.661457 & 0.345665 & 3.40175 & 0.328081 & 0.644351 & 0.083524 \\
\hline
0.60 & 0.433781 & 0.638470 & 0.337330 & 3.44352 & 0.335618 & 0.667248 & 0.084406 \\
\hline
0.56 & 0.414386 & 0.614513 & 0.327493 & 3.49485 & 0.344796 & 0.695714 & 0.085441 \\
\hline
0.52 & 0.393944 & 0.589494 & 0.315989 & 3.5579 & 0.355941 & 0.731158 & 0.086639 \\
\hline
0.48 & 0.372384 & 0.563306 & 0.302642 & 3.63551 & 0.369473 & 0.77551 & 0.088013 \\
\hline
0.44 & 0.349632 & 0.535823 & 0.287267 & 3.73153 & 0.385940 & 0.831469 & 0.089570 \\
\hline
0.40 & 0.325609 & 0.506896 & 0.269678 & 3.8513 & 0.406073 & 0.902923 & 0.091312 \\
\hline
0.36 & 0.300232 & 0.476343 & 0.249693 & 4.00246 & 0.430872 & 0.995668 & 0.093229 \\
\hline
0.32 & 0.273419 & 0.443942 & 0.227153 & 4.19634 & 0.461750 & 1.11871 & 0.095294 \\
\hline
0.28 & 0.245083 & 0.409409 & 0.201946 & 4.45054 & 0.500777 & 1.28676 & 0.097446 \\
\hline
0.24 & 0.215143 & 0.372374 & 0.174052 & 4.79392 & 0.551140 & 1.52543 & 0.099564 \\
\hline
0.20 & 0.183524 & 0.332334 & 0.143610 & 5.27761 & 0.618069 & 1.88328 & 0.101421 \\
\hline
0.16 & 0.150166 & 0.288556 & 0.111044 & 6.00181 & 0.710927 & 2.46347 & 0.102582 \\
\hline
0.12 & 0.115038 & 0.239887 & 0.077281 & 7.19438 & 0.848770 & 3.52552 & 0.102171 \\
\hline
0.08 & 0.078166 & 0.267729 & 0.044168 & 9.51648 & 1.079302 & 5.93005 & 0.098219 \\
\hline
0.04 & 0.039688 & 0.116656 & 0.015415 & 16.1086 & 1.58276 & 14.7202 & 0.085092 \\
\hline
0 & 0 & 0 & 0 & $\infty$ & $\infty$ & $\infty$ & $ $ \\
\hline
\end{tabular}
\end{table*}

\begin{table*}[b]
\centering
\caption{\label{tab:table2}%
Properties of the Maclaurin spheroids. The columns from left to right include the following dimensionless quantities: the eccentricity, $e^2=1-(a_3/a_1)^2$, where $a_1$ and $a_3$ are the semi-major axes with rotation along the $a_3$ axis, the dimensionless square of the angular velocity, the dimensionless rotation period, the dimensionless angular momentum, the dimensionless moment of inertia, and the dimensionless rotational kinetic energy.}
\begin{tabular}{cccccc}
\hline
$e$ & $\Omega^2/(\pi G \rho)$ & $P/\left(\pi G^{-1} \rho^{-1}\right)^{\frac{1}{2}}$ & $L/\left(G M^3 \bar{a}\right)^{\frac{1}{2}}$ & $I/ \left( M \bar{a}^2 \right)$ & $E_{\rm k}/\left(G M^2 \bar{a}^{-1}\right)$ \\
\hline
0 & 0 & $\infty$ & 0 & $ $ & 0 \\
\hline
0.10 & 0.00534 & 27.369 & 0.02539 & 0.4012 & 0.000803 \\
\hline
0.15 & 0.01204 & 18.2271 & 0.03829 & 0.402941 & 0.001819 \\
\hline
0.20 & 0.02146 & 13.6526 & 0.05144 & 0.405467 & 0.003263 \\
\hline
0.25 & 0.03363 & 10.906 & 0.06491 & 0.408712 & 0.005154 \\
\hline
0.30 & 0.04862 & 9.07032 & 0.07882 & 0.412761 & 0.007526 \\
\hline
0.35 & 0.06647 & 7.75742 & 0.09329 & 0.417822 & 0.010415 \\
\hline 
0.40 & 0.08727 & 6.77014 & 0.10846 & 0.423942 & 0.013874 \\
\hline
0.45 & 0.11108 & 6.00084 & 0.12450 & 0.431341 & 0.017968 \\
\hline
0.50 & 0.13799 & 5.38401 & 0.14163 & 0.440251 & 0.022781 \\
\hline
0.55 & 0.16807 & 4.87848 & 0.16013 & 0.451021 & 0.028426 \\
\hline
0.60 & 0.20135 & 4.45712 & 0.18037 & 0.464149 & 0.035046 \\
\hline
0.65 & 0.23783 & 4.10107 & 0.20286 & 0.480322 & 0.042838 \\
\hline
0.70 & 0.27734 & 3.79773 & 0.22834 & 0.500663 & 0.052070 \\
\hline
0.75 & 0.31947 & 3.53847 & 0.25792 & 0.526914 & 0.063125 \\
\hline
0.80 & 0.36316 & 3.3188 & 0.29345 & 0.562282 & 0.076574 \\
\hline
0.81 & 0.37190 & 3.27957 & 0.30153 & 0.570935 & 0.079624 \\
\hline
0.81276 & 0.37423 & 3.26934 & 0.30375 & 0.573345 & 0.080461\\
\hline
\end{tabular}
\end{table*}

\bibliography{sample631}{}
\bibliographystyle{aasjournal}

\end{CJK*}
\end{document}